# Structure and mechanical behavior of ultrafine-grained aluminum-iron alloy stabilized by nanoscaled intermetallic particles

Amandine Duchaussoy [a*], Xavier Sauvage [a], Kaveh Edalati [b,c], Zenji Horita [b,c], Gilles Renou [d], Alexis Deschamps [d], Frédéric De Geuser [d]

[a] *Normandie Univ, UNIROUEN, INSA Rouen, CNRS, Groupe de Physique des Matériaux, 76000 Rouen, France*
[b] *Department of Materials Science and Engineering, Faculty of Engineering, Kyushu University, Fukuoka 819-0395, Japan*
[c] *WPI, International Institute for Carbon-Neutral Energy Research (I2CNER), Kyushu University, Fukuoka 819-0395, Japan*
[d] *Univ. Grenoble Alpes, CNRS, Grenoble INP, SIMaP, F-38000 Grenoble, France*

**Abstract**

Ultrafine-grained aluminum alloys offer interesting multifunctional properties with a combination of high strength, low electrical resistivity, and low density. However, due to thermally induced grain coarsening, they typically suffer from an intrinsic poor thermal stability. To overcome this drawback, an Al-2%Fe alloy has been selected because of the low solubility of Fe in Al and their highly positive enthalpy of mixing leading to the formation of stable intermetallic particles. The two-phase alloy has been processed by severe plastic deformation to achieve simultaneously submicrometer Al grains and a uniform distribution of nanoscaled intermetallic particles. The influence of the level of deformation on the microstructure has been investigated thanks to transmission electron microscopy and atom probe tomography and it is shown that for the highest strain a partial dissolution of the metastable $Al_6Fe$ particle occurred leading to the formation of a Fe super saturated solid solution. The thermal stability, and especially the precipitation of particles from the ultrafine-grained solid solution and the way they pin grain boundaries has been investigated both from static annealing and *in-situ* transmission electron microscopy experiments. The correlation between microstructural features and microhardness has been established to identify the various strengthening contributions. Finally, it is shown that ultrafine grained high purity Al with less than 0.01 at. % Fe in solid solution could preserve a grain size only 300nm after 1h at 250°C.

1. Introduction

Aluminum alloys are attractive for applications where a combination of good electrical conductivity and low density is required. To increase the mechanical strength, precipitation hardening alloys can be used, yet with a degradation of the electrical conductivity. Another attractive strategy is to rely on grain boundary strengthening and to develop ultrafine-grained (UFG) structures. It has been shown that severe plastic deformation (SPD) such as achieved by equal-channel angular pressing (ECAP) or high-pressure torsion (HPT) may efficiently refine coarse-grained microstructures down to the sub-micrometer scale in a large variety of aluminum alloys [1–4]. It gives rise to high mechanical strength, potentially in combination with other properties such as electrical conductivity [2,5–10]. However, the extreme level of strain applied during these processes results in a very high level of stored energy that usually gives rise to a poor thermal stability of the microstructure and thus of the properties. Besides, a faster precipitation kinetics is also often reported in SPD precipitation hardening alloys [11,12]. It is attributed to a fastest atomic mobility promoted by crystalline defects and to a lower nucleation barrier on crystalline defects.

Several approaches have been proposed to stabilize UFG structures [13–16]. First, the free energy of grain boundaries can be reduced by solute segregation, thereby reducing the driving force for grain growth [13–15]. An example is segregation of Nd at grain boundaries of a QE22 Mg alloy as studied experimentally by Khan and Panigrahi [17]. Second, the mobility of grain boundaries can be reduced by solute drag effect [18] or by precipitates using the Zener pinning effect [19,20]. When appropriate combinations of size, number, density and volume fractions of precipitates are achieved, extreme thermal stability can be reached, such as in oxide dispersion strengthened steels where sub-micrometer grains can be stable up to 1100°C [21]. The critical grain size at which the grain growth stops due to Zener pinning is proportional to the particle size and inversely proportional to the particle volume fraction [22,23], thus a high volume fraction of uniformly distributed nanoscaled particles is required to inhibit grain growth. An additional requirement is the thermal stability of the pinning particles themselves. If they contain elements with a high diffusivity in the matrix (such as Mg, Cu, Si or Zn in Al matrix) then they will quickly coarsen leading to a reduction of the pinning point number density during thermal exposure. This phenomenon would be even faster if the solubility of the considered elements in the matrix is significant because of moving grain boundaries acting as "solute collectors"[24]. Therefore, an efficient Zener pinning of grains requires a high density of particles resistant to coarsening, which can be achieved best by phases containing elements of low diffusivity and low solubility.

In commercial aluminum alloys, such phases, containing elements of low diffusivity and low solubility, are typically created during the early stages of material processing by casting in the form of coarse, micrometer-scale particles containing Fe, Cr, and Mn. However, controlling their size and dispersion by a precipitation treatment is impossible since they are stable up to the liquidus temperature. However, when subjected to SPD, these brittle intermetallic particles can be fragmented and dispersed [25–28] or could even be dissolved within the matrix [5,7,26–31].

In this study, we applied this strategy of introducing a slow diffusing, intermetallic forming element, namely Fe, in Aluminum, in order to reach by SPD a finely dispersed distribution of thermally stable particles and evaluate how this distribution can improve the thermal stability of SPD-formed UFG structures. Fe is known to have a very low miscibility in Al [32], to form brittle intermetallics [33] and to have one of the lowest diffusivities in Al [34,35]. Besides, due to its low solubility, it can potentially help to reach interesting combinations of strength and conductivity as proposed by several authors [6,27,28]. Indeed, previous studies on various Al-Fe alloys deformed by severe plastic deformation have shown that grain refinement down to 400 – 100 nm is typically achieved leading to a strength between 250 - 625 MPa [29,36]. Intermetallic particles ($Al_6Fe$ and/or $Al_{13}Fe_4$) are fragmented by plastic strain and for the highest levels of deformation investigated, some iron in supersaturated solid solution with a concentration up to 2.2 wt. % has been reported [28]. The aim of the present work was to investigate at first the specific mechanisms leading to the UFG structure formation in Al-Fe alloys with a special emphasis on the progressive fragmentation of intermetallic particles, dissolution, and redistribution of Fe atoms. Then, the microstructure evolution during aging has been studied, particularly the precipitation of nanoscaled Fe rich particles within the UFG structure and their influence on grain boundary (GB) mobility and grain growth. To achieve these goals a combination of atomic-scale characterization by Atom Probe Tomography and Transmission Electron Microscopy with *in-situ* heating experiments has been carried out on an Al-2wt%Fe alloy deformed by HPT at various strain levels up to 1000 revolutions. The obtained microstructural data are used to estimate the different contributions (solid solution, GBs, dislocations, particles) to the strength evolution.

## 2. Experimental

The material investigated in the present study has been cast with a nominal composition of 2 wt. % Fe. The exact alloy composition (wt.%) is : Fe 1.99%, Si<0.004%, Mn 0.012%, Mg<0.189%, Cr 0.003%, Zn 0.002%, Ti<0.001% and Ni<0.041%. 10 mm discs with a thickness of 0.8±0.1 mm were cut from the ingot for HPT processing at room temperature, under a pressure of 6 GPa and with a rotational speed of 1 rpm. Different levels of deformation have been achieved by applying 10, 100 and 1000 revolutions. The anvil temperature, monitored during processing with an infra-red temperature measurement device, never exceeded 50°C for this material [37]. Static aging treatments were carried out in a furnace with an air atmosphere for one hour at 150°C, 200°C or 250°C. After aging, samples were quenched in iced water.

Microhardness in the initial states, after HPT processing and after aging were measured using a Vickers indenter on a Futur Tech FM-7e device, applying a load of 300g for a duration of 10 seconds, with 0.5mm steps along the disc diameter.

Microstructures were observed using several complementary techniques. Scanning electron microscopy (SEM) was performed with a Gemini Leo1530 FEG-SEM microscope using QBSD detector. Images were processed using ImageJ software. Transmission electron microscopy (TEM) was performed using a JEOL 2100F microscope equipped with automated crystal orientation mapping (ASTAR) [38–40]. Each grain is delimited from grain boundaries with minimum misorientation = 15°, their area is then measured and assimilated to a circle whose diameter is considered as the grain size. Additional data were obtained by Scanning TEM (STEM) using a JEOL ARM-200F operated at 200kV. Dark Field (DF - collection angles 20 to 80mrad) and high-angle annular dark field (HAADF – collection angles 80 to 180mrad) images were recorded with a probe size of 0.2 nm and a convergence angle of 34 mrad. Particles size was measured using HAADF images and the ImageJ software. Elemental analysis was carried out using Energy Dispersive X-ray Spectroscopy (EDS) with an Oxford Instruments X-max detector, which has a solid angle of 0.7 sr. *In-situ* TEM experiments were carried out with a double tilt heating holder (Gatan 652 MA) applying the temperature ramp depicted in Fig. 1. TEM samples were prepared as 3mm diameter discs whose center was halfway between the center and the edge of the HPT disc. At this location, the corresponding shear strain can be estimated by [41] :

$$\gamma = \frac{2\,\pi\,N.\,r}{h}$$

(1)

where $N$ is the number of revolutions by HPT, $r$=2.5mm the distance from the disc center and $h$=0.8mm the HPT disc thickness. Thus, samples observed after 10, 100 and 1000 revolutions correspond to an estimated shear strain of 300, 3000, 55000 respectively. For TEM observations, electron transparency was obtained by electropolishing (performed on a Stuers TenuPol-5) with a mixture of 30% $HNO_3$ + 70% $CH_3OH$ at a temperature of -30°C.

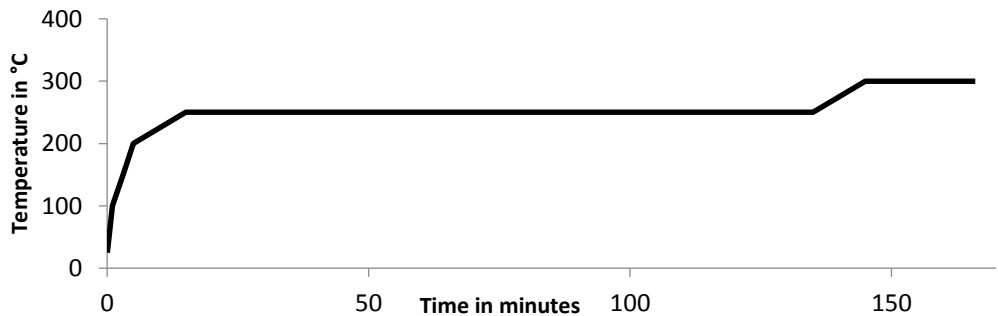

Fig. 1. Temperature ramp applied for the in-situ TEM experiment

Atom probe tomography (APT) analyses were carried out with a CAMECA LEAP-4000HR apparatus with a specimen temperature of 40K (-233°C), in ultra-high vacuum ($10^{-11}$ mbar), with a pulse fraction of 20% and a pulse repetition rate of 200Khz. 3D reconstructions were performed using IVAS Software and further data processing was done with the Gpm3dSoft software. APT tips were prepared using electro-polishing in a mixture of 75% $CH_3COOH$ + 25% $HClO_4$, and a mixture of 98% $C_8H_{16}O_3$ + 2% $HClO_4$ for the final part of the electropolishing, at the same location like TEM foils. For quantitative composition measurements, an isotopic overlap correction was necessary for the $^{27}Al^+$ and $^{54}Fe^{2+}$. This correction was based on the natural abundance of Fe. $^{57}Fe^{2+}$ is indeed not subjected to any overlap and the number of atoms detected in the corresponding mass range can be used to estimate the amount of $^{54}Fe^{2+}$ from the natural abundance isotopic ratio of Fe ($^{54}Fe$ : 5.845%, $^{57}Fe$ : 2.119%). Then, the number of $^{27}Al^+$ detected ions is estimated by subtraction [42]. The same method was applied to the overlap between $^{27}AlH^+$ and $^{56}Fe^{2+}$.

## 3. Results

### 3.1. As-cast microstructure

Fig. 2 shows a series of observations from the as-cast material microstructure used in this study. Fig. 2(a) shows the typical as-cast dendritic structure with pro-eutectic zones in dark (zone A) and Fe-rich eutectic zones in bright (zone B). Similar as-cast structures have been described in previous works [5–7,29,43–46]. The HAADF STEM image (Fig. 2(b)) shows similar zones at a higher magnification and clearly exhibits in bright the fine-scaled intermetallic particles within the eutectic areas (labeled B and B'). These particles are cylinder-shaped and depending on their orientation in the TEM thin foil they appear

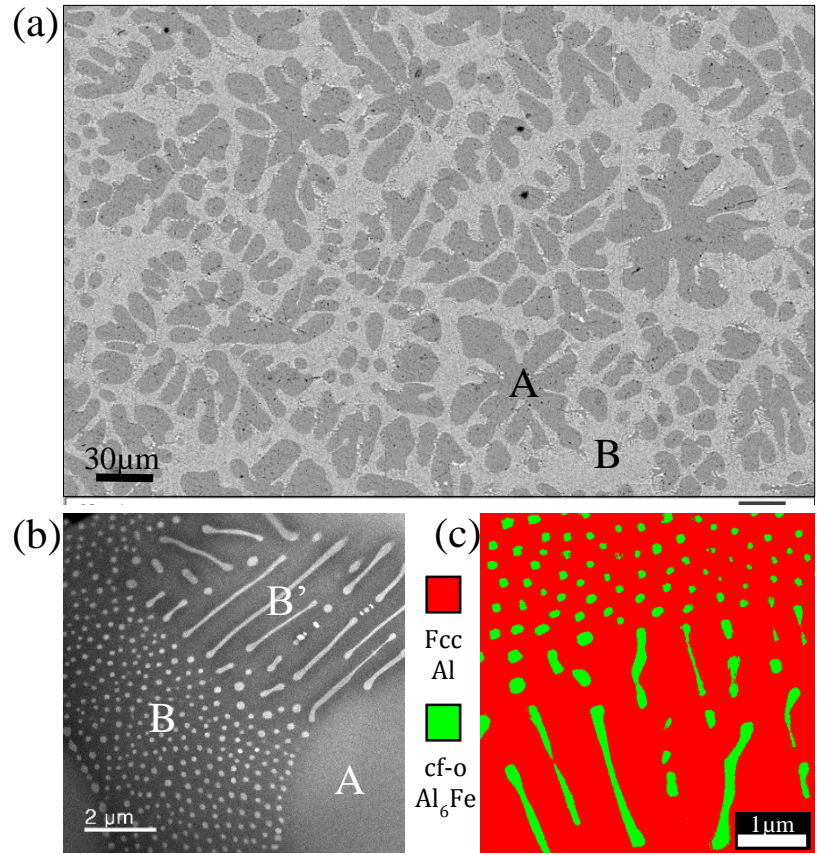

Fig. 2. (a) SEM micrograph (back scattered electrons detector); (b) STEM HAADF image showing different zones: Pro-eutectic zone (A) and eutectic zone (B and B'); (c) ASTAR phase map showing that the Fe rich intermetallic particles in the eutectic zones are $Al_6Fe$

equiaxed (zone B) or elongated (zone B'). Similar morphologies are exhibited on the phase map obtained with the ASTAR system (Fig. 2(c)). This map also confirms that the intermetallic particles embedded in the FCC Al matrix are metastable c-face centered orthorhombic $Al_6Fe$ phases in agreement with earlier works [5–7,29,43–46]. This phase exhibits a c-face centered orthorhombic crystal structure (Ccmm space group) with: a=6.49Å, b=7.44Å and c=8.79Å.

The composition of intermetallic particles measured by EDS analyses also confirmed that the Fe content is consistent with the $Al_6Fe$ stoichiometry (see Table. 1). The volume fraction of eutectic regions and of the intermetallic phase in the eutectic areas has been estimated by image processing of SEM and TEM images. By assuming that only Al$_6$Fe is present, a composition of 1.93±0.42 wt. % Fe is found, close to the nominal. Data are summarized in Table. 1.

| Volume fraction of eutectic zones | Volume fraction of intermatellic particles in eutectic zones | at.% Fe measured in intermetallic particles | Estimated wt.% Fe | Nominal wt.% Fe |
|---|---|---|---|---|
| **51 ± 2** | 14 ± 1 | 13.22 ± 3 | 1.93 ± 0.42 | 1.99 |

Table. 1. Mean volume fraction of eutectic in the as cast material estimated from SEM image processing (two images 256x389µm$^2$), mean volume fraction of intermetallic particles in eutectic areas estimated from STEM-HAADF image processing (7 images 6x6µm$^2$ and 3x3µm$^2$), mean at.%Fe in intermetallic particles measured by TEM-EDS and APT analyses, corresponding wt.% Fe estimated from previous values and nominal composition of the alloy (for comparison).

*3.2.    Impact of microstructural changes on microhardness*

Microhardness measurements have been used to evaluate the impact of the deformation by HPT and of the different aging treatments on the mechanical response. Fig. 3 shows the microhardness of as-cast and HPT discs plotted as a function of the distance to the HPT disc center and Fig. 4(a, b, c) shows the microhardness of HPT discs plotted as a function of the equivalent deformation.

As shown in Fig. 3, the microhardness of the as-cast sample is 41±3 HV, it is homogenous across the HPT disc diameter. After 10 revolutions by HPT (yellow curve), the microhardness is significantly higher but a strong gradient is exhibited from the disc center (72±7 HV) to the edge (145±9 HV). This behavior is a common feature of the HPT process for moderate de formation [47–52] and it results from the shear gradient along the radius induced by the torsion.

After 100 revolutions (green curve), this gradient is less pronounced and a saturation near a value of 173±18 HV is exhibited except in the central region up to 0.5 mm from the center. A fully homogeneous hardness is reached after 1000 revolutions (black curve) with a hardness value averaged on the whole sample of about 200±22 HV. Thus, even after 100 revolutions, the saturation in hardness has not been reached, indicating that some microstructural changes are still occurring up to 1000 revolutions.

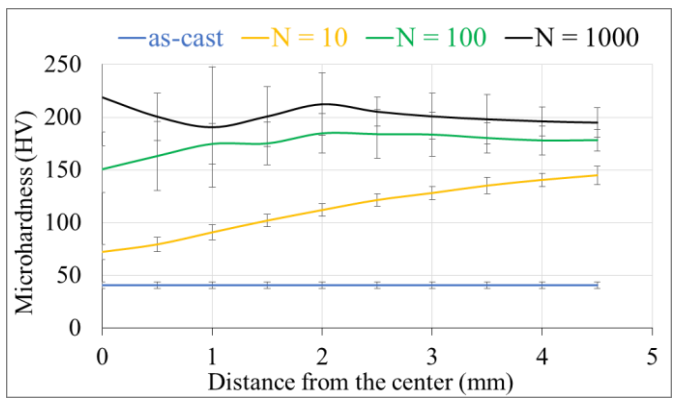

**Fig. 3. Evolution of the microhardness (HV) before and after HPT process.**

In Fig. 4, aging during 1h at 150°C or 200°C lead to a small HV increase of the material deformed by $N$=10 or $N$=100 revolutions (Fig. 4 (a,b)), while it remains constant for $N$=1000 (Fig. 4(c)). After aging at 250°C however, a significant drop is observed for all levels of deformation (for example from about 200 HV down to 100 HV for $N$=1000, see Fig. 4(c)) and initial gradients are strongly reduced.

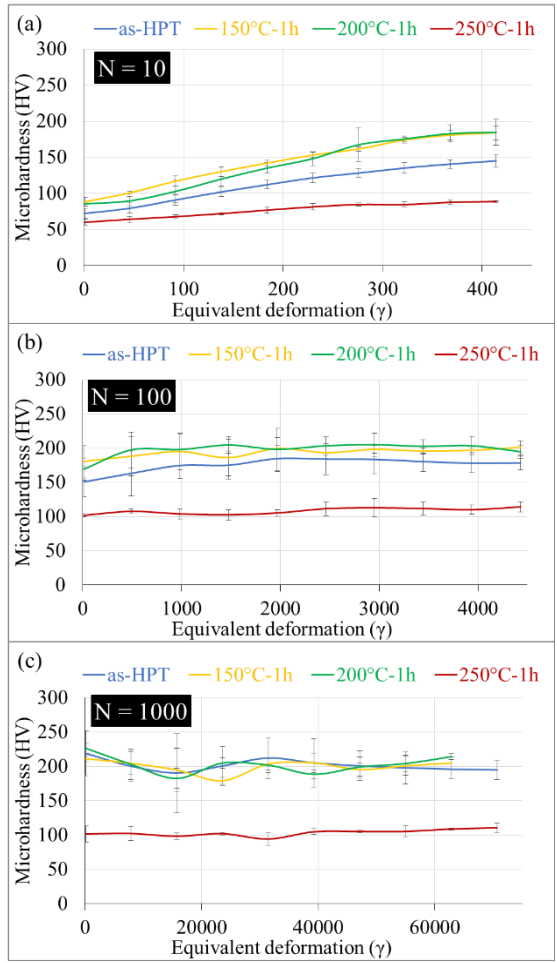

**Fig. 4. Evolution of the microhardness (HV) after aging at different temperatures for the different state of deformation: (a) N = 10 revolutions, (b) N = 100 revolutions and (c) N = 1000 revolutions**

*3.3.     Influence of severe plastic strain on the microstructure*

Ultrafine grains with a mean size well below the micrometer scale are clearly exhibited on STEM-DF images after *N*=10, 100 or 1000 revolutions (Fig. 5(a, b, c)). The grain size does not reach a steady state after 10 revolutions and further refinement is observed after 100 revolutions. At higher strain (*N*=1000), grains are not significantly refined but their shape changes from equiaxed to elongated, with an aspect ratio in a range of two to three.

The grain size distributions are plotted in Fig. 6(a), the mean grain size is about 300 nm for $N$=10 revolutions, down to 150 nm for $N$=100 and 140 nm for $N$=1000.

The influence of the strain level on the fragmentation of Fe-rich intermetallic particles can be seen on STEM-HAADF images thanks to the Z contrast in Fig. 5(d, e, f, g, h and i). After 10 revolutions, most of the particles exhibit a size that compares with the original $Al_6Fe$ cylinder diameter (mean size of 130 nm, see Fig. 5(d)). However, at a higher magnification, some fragmentation that occurred at a lower scale can be observed (arrowed in Fig. 5(g)). After 100 revolutions, some relatively large particles remain (Fig. 5(e)) but the large strain obviously led to extensive fragmentation and a high density of nanoscaled particles (Fig. 5(h)) with a mean size of about 70 nm. These small particles resulting from the fragmentation process are, at this stage, not homogeneously distributed in the Al matrix. After 1000 revolutions, further fragmentation dramatically decreased the proportion of large particles (Fig. 5(f)) and led to a relatively uniform distribution of nanoscaled size particles with a diameter down to 40 nm (Fig. 5(i)). STEM-HAADF images did not show any significant contrast enhancement at GBs that could be attributed to a strong segregation of Fe at GBs.

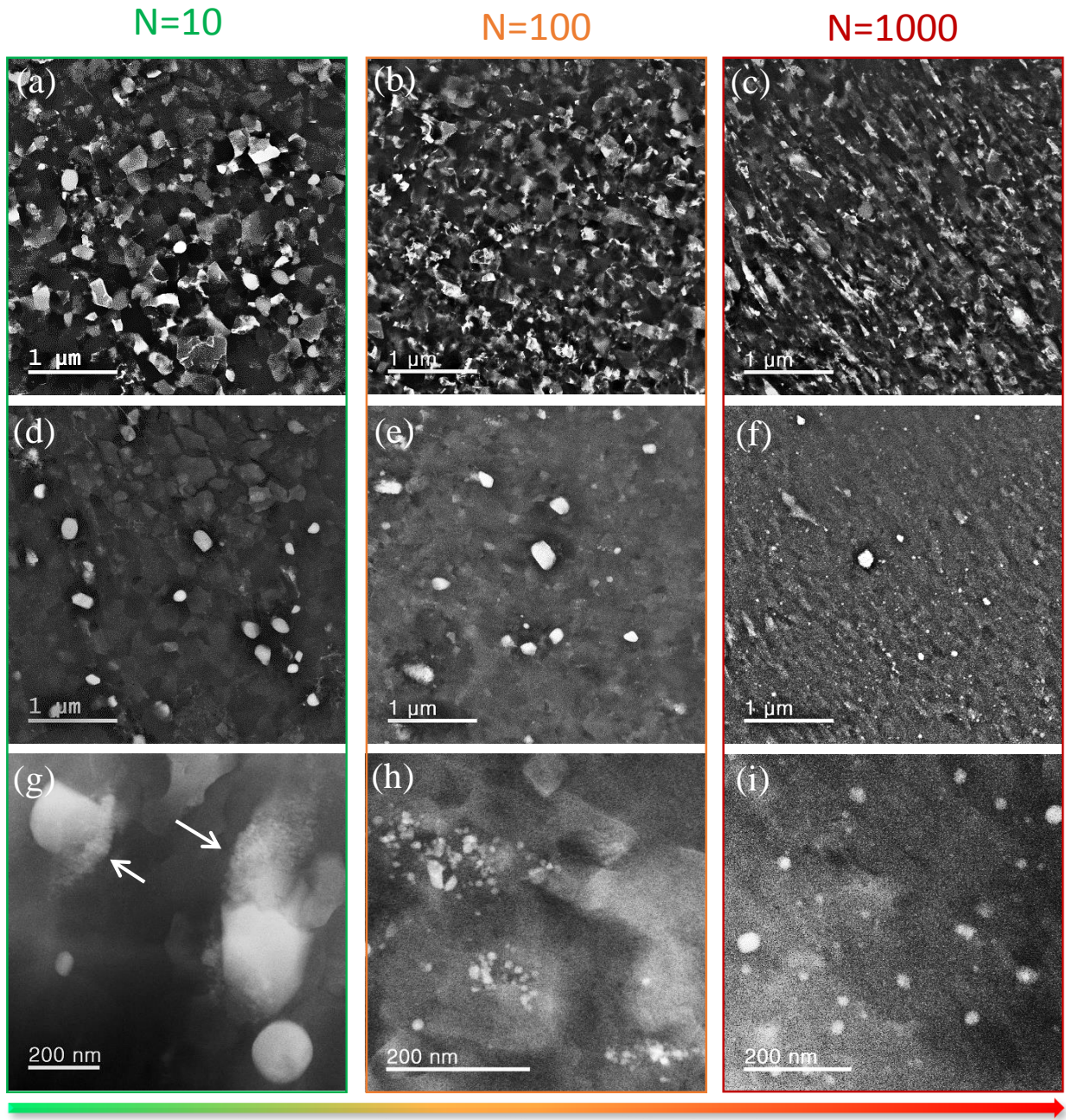

**Fig. 5.** STEM Dark Field (a, b, c) and STEM HAADF (d, e, f, g, h, i) images after HPT with N = 10 revolutions (a, d, g), N = 100 revolutions (b, e, h) and N = 1000 revolutions (c, f, i)

The evolution of the particle size distribution and of the relative volume fraction measured from STEM-HAADF images are plotted on Fig. 6(b) and (c). It clearly indicates that a huge strain (1000 revolutions, corresponding to a shear strain of 55000) is necessary to achieve a near complete fragmentation of the original coarse intermetallics to the nanoscale with a nearly uniform distribution of final particles. Electron diffraction in the TEM and TEM-EDS was applied on a large number of nanoscaled particles resulting from the fragmentation process (see supplementary material), and they all exhibited the crystallographic structure of the $Al_6Fe$ phase, indicating that no significant strain-induced phase transformation occurred during the HPT process.

It has been shown in the literature that intermetallic particles could be partly dissolved during SPD [5,6,26–30,53]. Measuring the volume fraction of intermetallic particles from STEM-HAADF images was only possible after 1000 revolutions where particles are homogeneously distributed and where it is possible to obtain representative data from a limited number of images. Assuming that nanoscaled particles are spherical and using the distribution plot of Fig. 6(b) the volume fraction of $Al_6Fe$ phase after a strain of 55000 was estimated in a range of 1.43 % to 3.34 % (for a TEM thin foil thickness in a range of 30 to 70 nm). This is significantly lower than the original volume fraction of about 7.14±0.51 %, suggesting that some strain-induced dissolution occurred. Since the equilibrium solubility of Fe in fcc Al is extremely low (between 0.025-0.052 at. % for temperature between 655-660°C [32,54,55] and even lower at room temperature), some APT analyses have been performed to reveal the distribution of Fe atoms after 1000 revolutions by HPT.

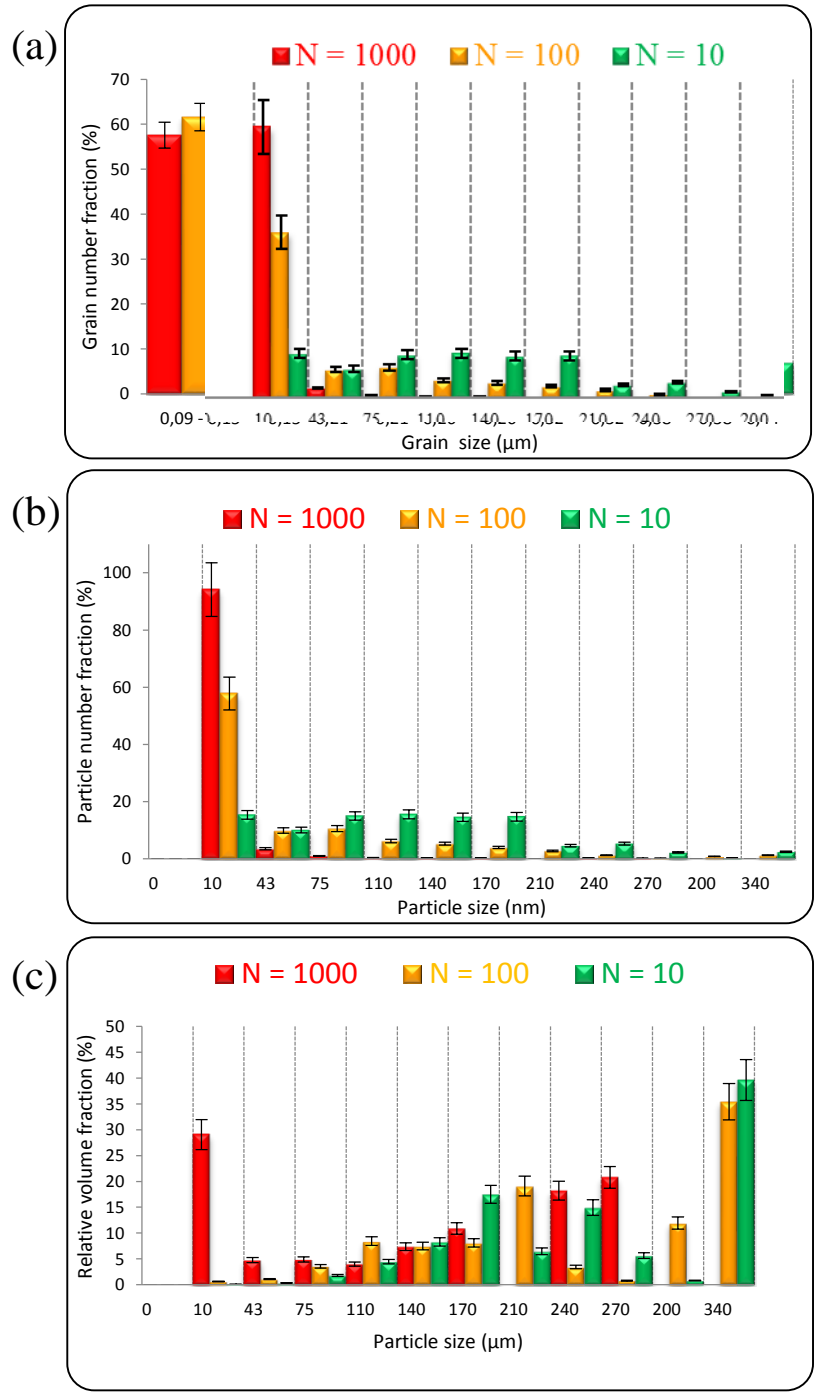

**Fig. 6.** Distribution of the grain size (a), the particle size (b) and the particle volume fraction (c) after HPT up to N = 10, 100 and 1000 revolutions

One of the analyzed volumes is displayed in Fig. 7. To clearly exhibit Fe-rich nanoscaled particles, the dataset has been filtered. Due to the relatively low Fe concentration of these particles and the overlap in the mass spectrum between the molecular ions $^{27}AlH^+$ and $^{56}Fe^{2+}$ ions (corresponding to the main Fe isotope), data have been filtered using a threshold on $Al^{2+}$ ions. Indeed, $Al_6Fe$ particles exhibit a higher evaporation field than the matrix, which leads locally to a stronger $Al^{2+}$ / $Al^+$ ratio than in the matrix (22 % against 12 % on average). The largest particles that are visible on the image are elongated with a mean size of about 14 nm.

The Fe concentration in the three particles that are exhibited in Fig. 7(a-b) is 11.45±0.95 at. %. These values are near the stoichiometry of the $Al_6Fe$ phase. Interestingly, the filtering procedure also reveals a local Fe enrichment (composition between 0<at.% Fe<8.8) that follows a line and which could be attributed to Fe segregation along a dislocation (see supplementary material S1 where the video showing the rotation 3D volume clearly show the linear feature). The Fe concentration in solid solution in the Al matrix (out of particles and segregation line) has been measured in various places and is ranging from 0.36 to 0.89 at. %. This is about 10 times the equilibrium solubility at 660°C, thus confirming the partial strain-induced dissolution of $Al_6Fe$. Based on TEM images, the mean volume fraction of $Al_6Fe$ phase was estimated as 7.14±0.51 % in the as-cast condition and ranging from 1.43 % to 3.34 % after 1000 revolutions. Thus, the dissolved volume fraction of $Al_6Fe$ particles during the deformation is 4.75±1.47 % corresponding to a Fe concentration in solid solution ranging from 0.47 to 0.89 at. %. These estimates are in good agreement with the range measured by APT (0.36 to 0.89 at%).

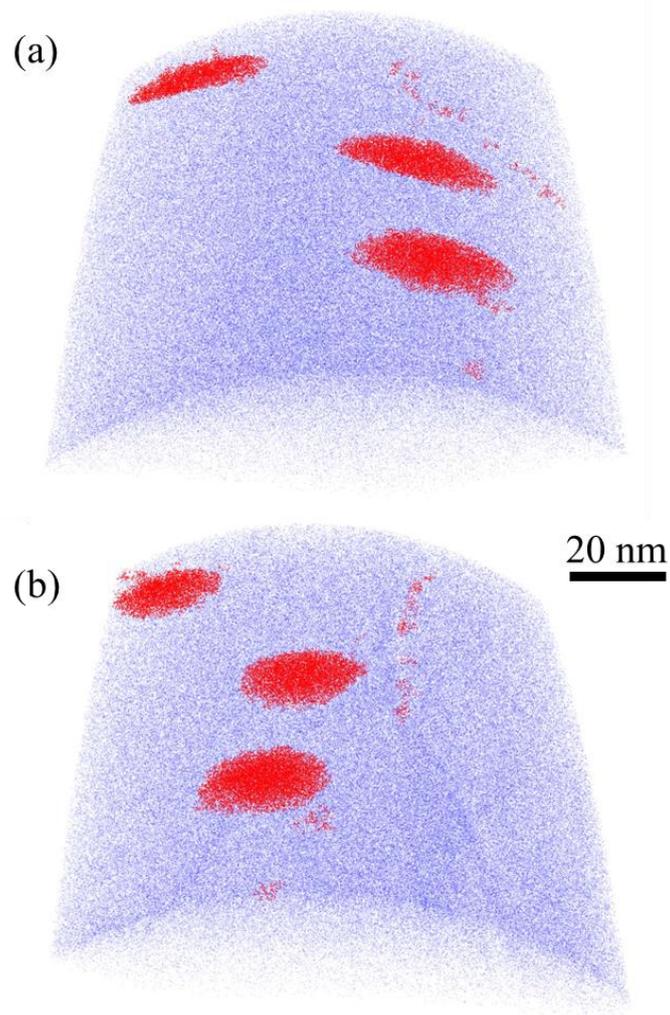

Fig. 7. (a) 3D reconstruction of a volume (160x160x105 nm$^3$) analyzed by APT in the N = 1000 HPT deformed sample. (b) Same volume tilted by 90° along vertical axis. The data have been filtered (see text for details) to display Fe rich areas (in red), blue dots are aluminum atoms. Three nanoscaled particles and a line of segregation of Fe are exhibited.

*3.4.    Evolution of the ultrafine-grain structure during aging*

*In-situ* heating experiments in the TEM have been carried out on the material deformed up to 1000 revolutions in order to follow the evolution of the UFG structure achieved by HPT, and especially to investigate the decomposition mechanisms of the Fe supersaturated solid solution and the influence of nanoscale intermetallic particles on grain boundary mobility. A progressive *in-situ* heating (Fig. 1) of the sample was performed up to 250°C and then the temperature was kept constant during 2h to follow the grain growth using STEM-DF and intermetallic particle nucleation and growth using STEM-HAADF. Then, the sample was further aged for 20 min at 300°C. The reader should note that DF images (Fig. 8(a, c, e)) are at relatively low magnification to highlight grains, while HAADF images (Fig. 8(b, d, f)) are at higher magnification to highlight nanoscaled particles within the grains or at GBs. The locations seen on initial images (Fig. 8(a, b)) and after 60 min at 250°C (Fig. 8(c, d)) slightly differ because of some unavoidable specimen drift. Few particles have however been circled to guide the eye. As shown in (Fig. 8(b, d, f)), some intermetallic particles grew (circled in green), others have nucleated at grain boundaries (circled in orange), while few others apparently do not significantly change (circled in blue). It is important to note that homogeneous nucleation of some Fe-rich particles inside Al grains has never been observed. Besides, even after 1h at 250°C, the grain size apparently does not change significantly, and grain growth seems to start only at 300°C (Fig. 8(a, c, e)).

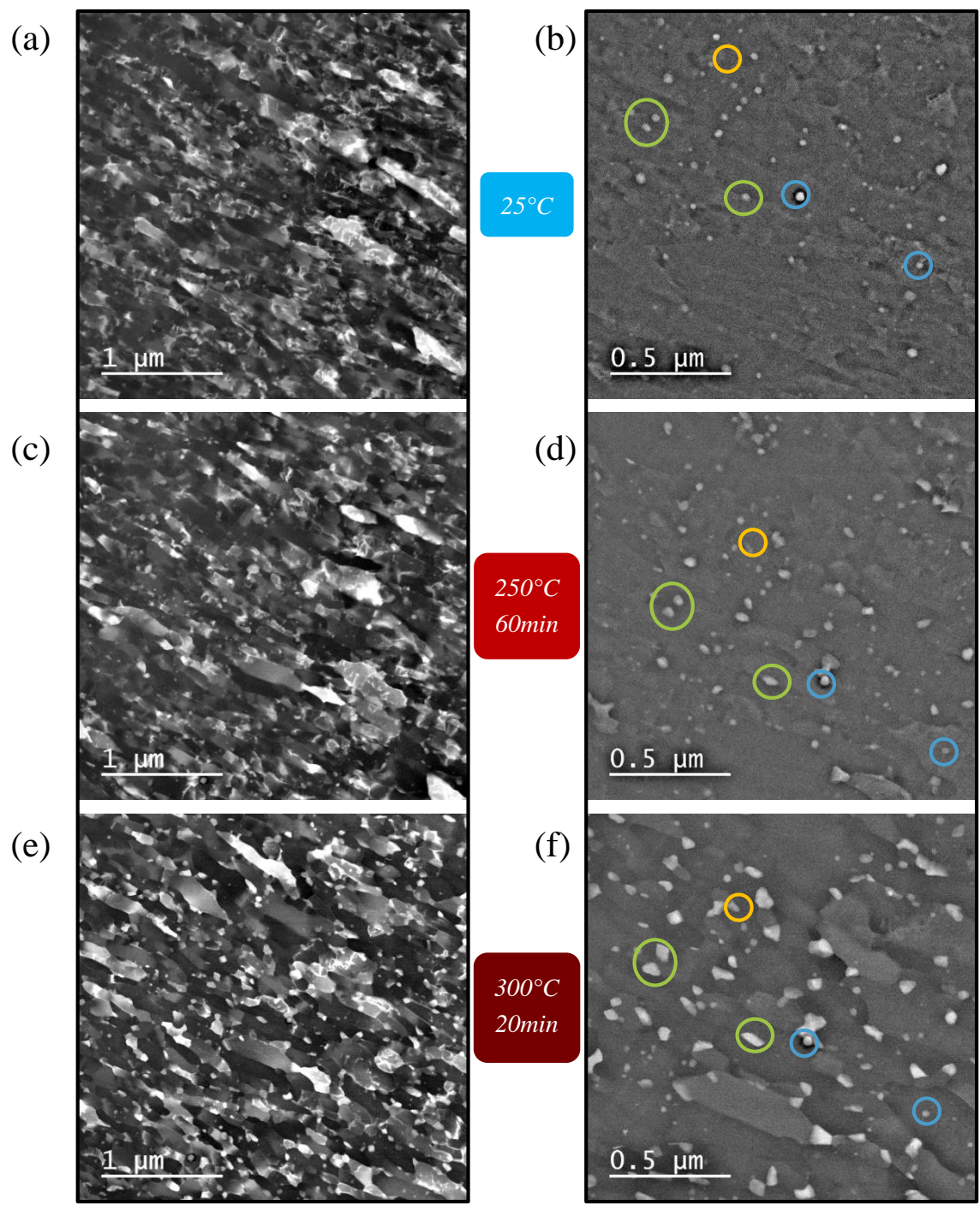

**Fig. 8. In-situ TEM on N = 1000 revolutions sample. Dark-field (a, c, e) and STEM-HAADF images (b, d, f) from 25°C (a, b), to 250°C during 60 min (c, d) and some additional 20 min at 300°C (e, f). HAADF images show the nucleation and growth of intermetallic particles (circled) and the dark-field images, the relatively stable grain size.**

In Fig. 9, the size of some particles is plotted as a function of aging time at 250°C, confirming the existence of different particle families: particles with very high growth rate (green curves), particles with a lower growth rate (orange curves) and particles which practically do not grow (blue curves).

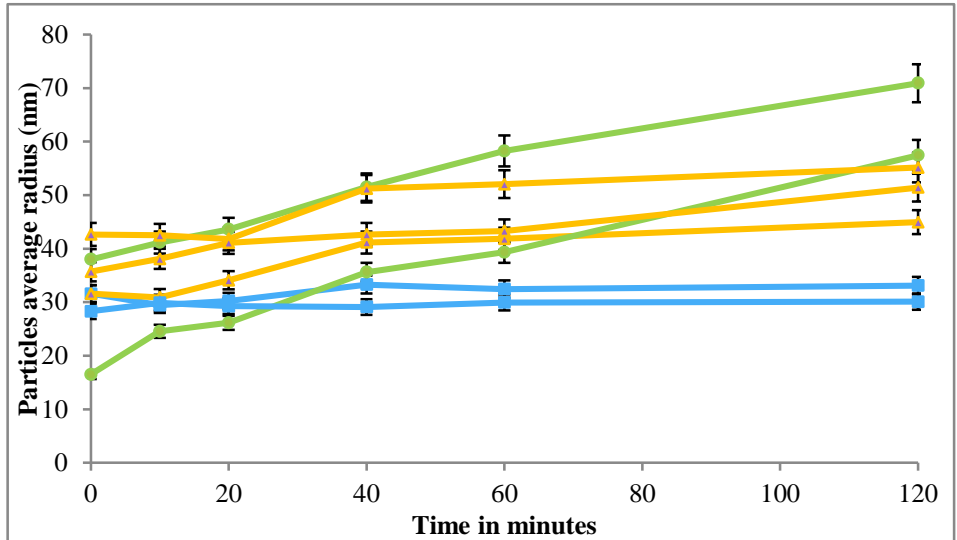

Fig. 9. Evolution of the size of some individual intermetallic particles during in-situ TEM heating at 250°C (each colored line corresponds to the growth of a specific type of particle (circled Fig.7.). Green for fast growing particles, orange for slow growing particles and blue for stable particles)

Because of possible artifacts related to the *in-situ* heating experiments (such as grain boundary pinning by the TEM thin foil surfaces), some static aging during 1h at 250°C has also been carried out. Fig. 10 shows TEM-ASTAR orientation maps, (a) after 1000 revolutions by HPT and (b) followed by 1h of static aging at 250°C. Some grain growth obviously occurred but the final grain size is still well below a micrometer. Besides, the phase map (Fig. 10(d)) clearly shows that two types of Fe-rich intermetallic particles are present within the microstructure after aging, namely $Al_6Fe$ and $Al_{13}Fe_4$. TEM- ASTAR and STEM-EDS performed on various particles also indicated that both Al/Fe ratios corresponding to these two phases could be detected (data not shown here). Our observations suggest that the new particles nucleated from the supersaturated solid solution are from the stable $Al_{13}Fe_4$ phase, while others are the initial nanoscale $Al_6Fe$ particles left after HPT and that have grown during aging.

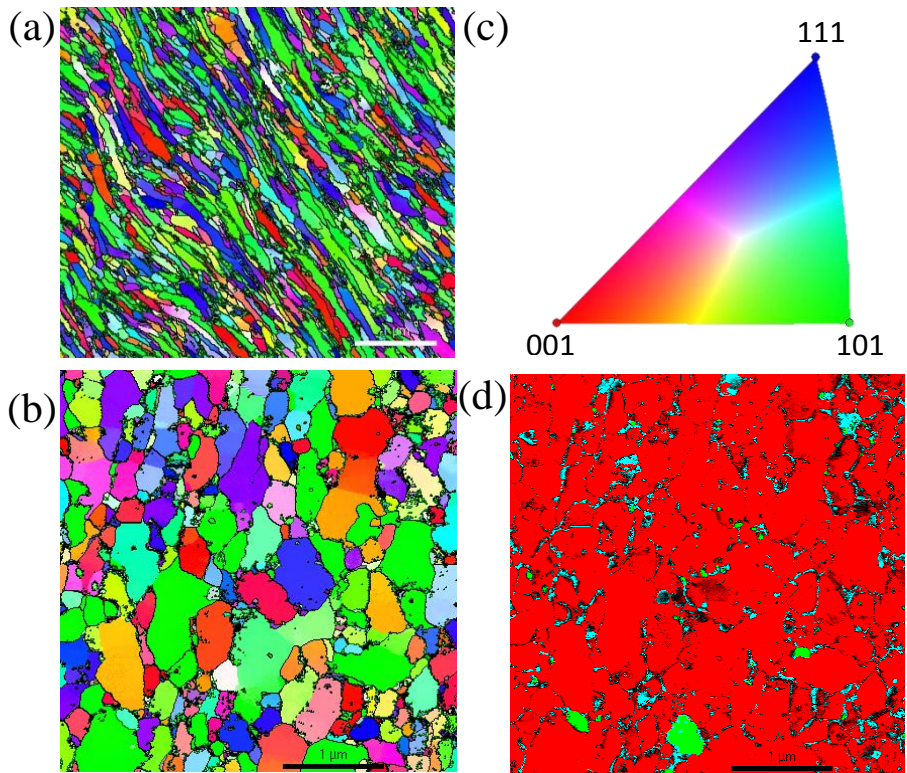

**Fig. 10.** (a) ASTAR orientation map showing the N = 1000 as-deformed microstructure. (b) ASTAR orientation map after static annealing at 250°C during 60min, (c) the orientation color code of images (a) and (b). (d) Phase map of the same area like (b) and showing both $Al_6Fe$ and $Al_{13}Fe_4$ phases are detected.

The evolution of the mean grain size, particle size and volume fraction during static aging of the different samples deformed by HPT are summarized in Table. 2. In this table, 100 and 1000 revolutions samples show the same mean particle size after aging but the total volume fraction depends on the respective proportion of $Al_6Fe$ and $Al_{13}Fe_4$ phases. In the material deformed up to *N*=10 revolutions, the mean size of intermetallic particles remains almost constant during aging because they are relatively coarse. However, in the material deformed by 100 revolutions, the mean size decreases because numerous new nanoscaled particles have nucleated out of the supersaturated solid solution shifting the distribution to a smaller size. It is interesting to note that this phenomenon does not affect the mean size in the sample deformed by 1000 revolutions because of the fast growth of particles in the early stage of deformation. Finally, the smallest grain size after static annealing is obtained for the highest level of deformation which provided the finest distribution of nanoscaled intermetallic particles that could pin GBs and inhibit grain growth (Table 2).

|  | N = 10 | | N = 100 | | N = 1000 | |
| --- | --- | --- | --- | --- | --- | --- |
|  | As-HPT | After Ageing (250° C/1 h) | As-HPT | After Ageing (250° C/1 h) | As-HPT | After Ageing (250° C/1 h) |
| D (nm) | 305±45 | 740±30 | 154±16 | 403±22 | 142±40 | 304±13 |
| d (nm) | 134±8 | 137±6 | 70±4 | 40±5 | 40±7 | 40±5 |
| $F_v$ % | 6.83 | 6.83 | 6.45±2.66 | 4.15-6.83 | 2.4±1.4 | 4.15-6.83 |
| at%Fe in solid solution | - | - | - | - | 0.6±3 | <0.01 |
| Dislocation densities ($\rho$ m$^{-2}$) | 8.16 x 10$^{14}$- 1.10 x10$^{15}$ | 3.99 – 4.31 x10$^{14}$ | 2.14 – 2.58 x10$^{15}$ | 1.28 – 3.0 x10$^{14}$ | 2.0 – 3.35 x10$^{15}$ | 8.32 x10$^{12}$- 2.73 x10$^{14}$ |

**Table. 2. Average grain size (D) estimated from ASTAR orientation images, average particle size (d) and volume fraction of particles (fv) estimated from STEM-HAADF images, at.% Fe in solid solution in the fcc Al matrix measured by APT, and dislocation densities**

To confirm that after aging the Al matrix has recovered a Fe content close to the thermodynamic equilibrium, APT measurements were carried out on the 1000 revolutions sample after static aging at 300°C during 3h. These data (3D volume not shown here) show that only a small residual fraction of Fe is still in solid solution after this aging treatment, the Fe content being close to the detection limit (lower than 0.01%at).

## 4. Discussion

### *4.1. Grain refinement, fragmentation, and dissolution of intermetallic particles induced by SPD*

The as-cast microstructure exhibits pro-eutectic FCC-Al and eutectic two-phase regions containing FCC-Al and the metastable $Al_6Fe$ phase (Fig. 2). In similar Al-Fe alloys, the stable $Al_{13}Fe_4$ phase is often reported in as-cast conditions [5–7,29,44–46] but it has also been shown that fast cooling conditions usually promote the formation of $Al_6Fe$ [43]. This metastable phase exhibits a higher Al / Fe ratio and thus a larger volume fraction is formed. This is beneficial for the approach proposed in this study where a fine distribution of intermetallic is targeted to stabilize the UFG structure.

After 10 revolutions by HPT, this initial as-cast structure is completely transformed into a relatively homogeneous microstructure with a mean grain size of about 300 nm which is consistent with typical data reported in the literature for similar or other Al alloys deformed by SPD in similar conditions [26,56]. At this stage, the particle fragmentation is not pronounced (mean particle size is decreasing from 240 nm (as cast) to 130 nm) but already affects the grain refinement of the Al matrix. Indeed, commercially pure Al deformed by HPT in similar conditions typically leads to a mean grain size of about 800 nm [49]. It has been shown that, in some cases, the presence of second phase particles can influence the grain refinement due to the local stress-induced next to particles [57]. In the present Al-2%Fe alloy, similar features probably occur. After 100 and 1000 revolutions, the mean grain size progressively decreases down to 150 nm and 140 nm respectively (Fig. 5). Although the grain size remains relatively unchanged between 100 and 1000 revolutions, the mean particle size continues to decrease (from 70 to 40 nm) and the spatial distribution of particles evolves toward a more homogenous distribution (Fig. 6).

The extremely high strain level also leads to a progressive strain-induced dissolution of intermetallic particles and the formation of a Fe supersaturated solid solution (Fig. 7 and table 2). Similar features have been reported for other immiscible systems [58–60], and also on Al-Fe alloys processed by SPD but based on indirect methods relying on XRD analyses [28–30,53]. In the present work, our direct measurements using APT prove that Fe is really in solid solution into the Al matrix and our data also indicates that there is a strong interaction between dislocations and Fe atoms. Then, one can propose that dislocations cutting through intermetallic particles play a significant role during the straining process. They could drag Fe atoms out of the particles and progressively dissolve them.

### *4.2. Precipitation and particle growth during post-deformation annealing*

The UFG structure achieved by HPT progressively transforms during low-temperature aging and the Fe supersaturated solid solution is progressively decomposed via several concomitant mechanisms (Fig. 8): the remaining $Al_6Fe$ particles keep the same crystallographic structure and grow, while new particles with the stable $Al_{13}Fe_4$ structure nucleate and grow on crystalline defects. These particles pin the GBs which limits the grain growth and after 1h at 250°C the mean grain size is only about 300nm in the Al-2%Fe alloy processed by HPT up to 1000 revolutions (Table 2).

Interestingly, it was found that the particle growth rate could be very different from one particle to another (Fig. 9) which is usually not the case for homogeneous nucleation and growth. In the literature, there is a large scatter in bulk diffusion data for Fe in Al [34,61–65]: the pre-exponential factor $D_0$ ranges from 91 to 5.3 $10^{-3}$ m$^2$/s and the activation energy Q ranges between 183.4 and 258.7 kJ/mol. To estimate the equivalent diffusion distance $X$ of Fe in Al the following equation was used [66,67]:

$$X = \sqrt{6Dt} \qquad (2)$$

Where D is the diffusion coefficient *(D=D$_0$ exp (-Q/RT), D$_0$ the frequency factor for Al, T the temperature, R the gas constant)* and *t* the time. It yields: 1.5 $10^{-4}$ < X < 0.1 nm for 1h at 150°C and 0.2 < X < 8.7 nm for 1h at 250°C. These values are obviously much smaller than the mean distance between particles or than the grain size. Thus, it indicates that the diffusion of Fe atoms is strongly promoted by crystalline defects like dislocations and GBs. Therefore, the growth rate of particles strongly depends on their location within the UFG structure (at GBs, at triple lines, along dislocations). Besides, some heterogeneities in the original Fe atom distribution may also affect the local growth rate. This relatively high diffusivity of iron atoms along GBs, promote the coarsening of particles and thus may affect the thermal stability through a reduction of the Zener pinning effect.

### *4.3. Stabilization of the UFG structure by nanoscaled intermetallic particles, Zener pinning*

During annealing, it has been observed that following the theory of Zener, nanoscaled particles pin GBs and limit the grain growth. Thus, using the Zener pinning model [18,20,68], the grain size $D_{lim}$ where an equilibrium between the growing pressure and the pinning pressure by the particles is reached can be estimated:

$$D_{lim} = \frac{2}{3}\frac{\Phi}{f} \qquad (3)$$

where ($\phi$) is the mean particle radius and $f$ is the particle volume fraction. Gladman [69] has modified this expression to consider the distribution of grain size by introducing a term $Z$ (ratio of growing grain size to average grain size):

$$D_{lim} = 0.41\frac{\Phi}{f} \qquad (4)$$

More recent works [70,71], have shown however that, in case of high-volume fraction of particles (between 0.01 and 0.1), others particle-boundary correlation factors have to be taken in account, leading to :

$$D_{lim} = \beta\frac{\Phi}{f^{1/3}} \qquad (5)$$

Where β is a geometric constant. As suggested by Hillert [70] and by Samuel A. Humphry-Baker and Christopher A. Schuh [71] β ranges from 1.8 to 1.6 respectively.

The equilibrium grain sizes for the as-deformed Al-2%Fe alloy have been estimated from eq(4) and eq(5), assuming that all intermetallic particles are from the $Al_6Fe$ phase (Table 3).

| Grain size in given in nm | N = 10 | N = 10 + 250°C-1h | N = 100 | N = 100 + 250°C-1h | N = 1000 | N = 1000 + 250°C-1h |
|---|---|---|---|---|---|---|
| Mesured | 305±45 | 740±30 | 154±16 | 403±22 | 142±40 | 304±13 |
| Calculated eq (4) (Gladman [69]) | 805±48 | 823±36 | 421±25 | 327±117 | 624±109 | 327±117 |
| Calculated eq (5) (β = 1.8 [70]) | 590±35 | 604±27 | 309±18 | 194±40 | 243±43 | 194±40 |
| Calculated eq (5) (β = 1.8 [71]) | 588±32 | 537±24 | 274±16 | 173±36 | 216±38 | 173±36 |

**Table. 3. Experimental (measured) and theoretical (from eq(4) and eq(5)) different limiting grain size (in nm) reached during thermal annealing for different states.**

Theoretical estimates of grain sizes for as-HPT microstructures are systematically larger than experimental values. This indicates that as-HPT microstructures are not in equilibrium with respect to grain growth. This is consistent with a grain refinement mechanism controlled by dynamic recovery processes during SPD at low homologous temperature (<0.5Tm).

After aging at 250°C during 1h, the volume fraction and size of nanoscaled intermetallic particles has changed and the measured grain size systematically lies between equilibrium grain sizes estimated from eq (4) and eq (5). One should note that the large error bar in estimated values at N = 100 and 1000 after annealing given by eq(4) is due to the proportion of $Al_{13}Fe_4$ which is difficult to estimate accurately from our observations. Eq(4) gives 210 nm if all intermetallic particles are considered as $Al_6Fe$ and 444 nm if they are all transformed or precipitated as $Al_{13}Fe_4$, that gives average of 327 nm.

Thus after aging 1h at 250°C one may consider that, the UFG structure has reached a relatively stable grain size. In any case, it is clear that further growth of intermetallic particles at this temperature cannot be avoided and will lead to a progressive (but slow) grain growth if the material is held at this temperature during a longer time.

*4.4.  Relationship between microstructural features and microhardness*

As expected [34,65], the large decrease in grain size comes together with an increase of hardness: from 45 HV for the as-cast sample to ~115 HV for the sample deformed to 10 revolutions, ~175 HV for the sample deformed to 100 revolutions and ~200 HV for the sample deformed to 1000 revolutions. However, this microhardness evolution is likely not only induced by grain refinement, since other obstacles to moving dislocations evolve with deformation: dislocation density, intermetallic particle number density, Fe in solid solution or segregated along dislocations [72].

The impact of iron in solid solution on strengthening can be estimated from the work by Mahon and Marshall [73]. Their results from low solute contents can be extrapolated with the Labusch law [74] resumed here with a 2/3 exponent.

$$\Delta\tau_{SS} = \frac{G[\varepsilon'G^2 + (15\varepsilon_b)^2]^{2/3}c^{2/3}}{\varphi} = A \times c^{2/3} \qquad (6)$$

where $G$ is the shear modulus, $\varepsilon'G$ the modulus of mismatch, $b$ the Al burgers vector magnitude, $\varepsilon_b$ the atomic size mismatch, $\varphi$ a constant and $c$ the concentration in solute element.

Intermetallic particles also contribute to the strengthening because the fragmentation leads to a progressive increase of the particle number density. Assuming that half of the Fe is in solid solution in the sample with 1000 revolutions (in accordance with APT measurements), the remaining Fe within Al$_6$Fe particles (of 40 nm diameter) lead to an intermetallic volume fraction of 2.6%. The corresponding Orowan strengthening contribution can be estimated by [75–77]:

$$\Delta\sigma = 0.7 M\mu b \frac{\sqrt{f_v}}{R} \qquad (7)$$

where $M = 3.06$ is the Taylor factor, $\mu = 26.9$ GPa the shear modulus, $b = 0.286$ nm the Burgers vector, $f_v$ the volume fraction and $R$ the radius of the particles.

The contribution of grain boundaries can be estimated through the Hall-Petch relationship [78,79]:

$$\Delta\sigma = \frac{k}{\sqrt{d}} \qquad (8)$$

with $0.07 < k < 0.09$ MPa.m$^{1/2}$ [80] and $d$ the average grain size estimated from TEM orientation mapping.

The contribution of the dislocation density can be estimated with the Bailey-Hirsh law [81,82]:

$$\Delta\sigma_{dislo} = \alpha M\mu b \rho^{\frac{1}{2}} \qquad (9)$$

where $\alpha$ is the strength factor for dislocation strengthening, with a typical value of 0.2 [83], and $\rho$ the dislocation density. The influence of possible Fe segregation along some dislocations (as revealed in Fig. 7) is not considered. Thus the contribution of Fe in solid solution is only estimated from eq(6).

The estimations of hardening contributions (done through the Tabor's law where $\sigma \approx 3$ HV) was done using experimental data reported in table 2 and are listed in table 4. Their contribution to the hardness has been estimated based on experimental hardness values (Table 4) by subtracting other contributions (including an intrinsic yield stress for Al of 10 MPa), and the dislocation density (Table 2) was then calculated based on eq(9). The dislocation densities estimated that way are consistent with dislocation densities typically reported for severely deformed pure Al ($1.8 \pm 0.3 \; 10^{14}$ m$^{-2}$ [76]), Al-1%Mg alloy ($3.9 \pm 0.4 \; 10^{14}$ m$^{-2}$ [76]), Al3%Mg alloy ($2.3 \pm 0.2 \; 10^{15}$ m$^{-2}$ [76]) or a 2024 Al alloy (~$8.9 \; 10^{14}$ m$^{-2}$ [75]) estimated by diffraction profile analysis.

Grain boundary and particle contributions increase with deformation because of grain size refinement and intermetallic particle fragmentation. The solid solution contribution cannot be estimated for $N=$ 10 and 100 revolutions because the microstructure is heterogeneous and if any Fe in solid solution it is probably located only in the vicinity of particles that started to split. However, it is interesting to note that the largest difference between estimated hardness contributions and measured values (reflected in HV$_{dis}$ in Table 4) is significantly larger for $N=100$ than for any other states (also giving rise to a higher estimated dislocation density in Table 2). This probably indicates that locally some Fe has already been inserted in solid solution and that it starts affecting the total hardness.

It is noted that during aging at 150°C and 200°C, a small but significant increase in hardness is observed (Fig. 4). Similar hardening by annealing has been reported by several authors in some metallic alloys processed by SPD [5,27,29,84,85]. This could be attributed to precipitation of second phase particle, but this is very unlikely in our case, especially in the material processed by $N$=10 revolutions where original intermetallic particles are still relatively coarse and where the amount of Fe driven in solid solution is very low. As Gammer and co-authors proposed [86], this could be attributed also to further grain size refinement induced by recovery mechanisms that convert low-angle GBs into large-angle GBs. However, in the present study, this phenomenon could not be evidenced experimentally. At last, other authors have proposed that such hardening by annealing might be attributed whether to dislocation starvation or relaxation of non-equilibrium GBs resulting from the SPD process [87]. It has not been however possible to confirm this hypothesis in the present investigation.

After aging at 250°C, the microhardness decreases to finally reach ~80 HV for $N$=10 revolutions and ~100 HV for $N$=100. This can be attributed to particles growth and grain growth. Indeed, the grain size increases from 300 to 740 nm for $N$=10 and from 150 to 400 nm for $N$=100. It is however interesting to note that a grain size of only 400 nm is kept after annealing at 250°C during one hour which is a remarkable thermal stability as compared to conventional alloys where extensive grain growth occur as early as 100-150°C [88,89].

|  | **10T** | **10T + 250°C-1h** | **100T** | **100T + 250°C-1h** | **1000T** | **1000T + 250°C-1h** |
|---|---|---|---|---|---|---|
| **HV$_{measured}$** | 115±30 | 80±11 | 173±18 | 109±11 | 200±22 | 104±12 |
| **HV$_{GB}$** | 45-52 | 30-32 | 65-72 | 41-43 | 63-84 | 47-50 |
| **HV$_{part}$** | 14 | 14 | 27 | 37-47 | 29 | 37-47 |
| **HV$_{ss}$** | 0 | 0 | 0 | 0 | 14-27 | 0 |
| **HV$_{dis}$** | 45-53 | 32-33 | 73-81 | 15-28 | 57-92 | 5-17 |
| **Hv$_{tot}$** | 107-122 | 79-82 | 168-183 | 96-121 | 166-235 | 92-117 |

Table. 4. Average micohardness measured at 4mm from the disc center (HV$_{measured}$) compared to estimations of the different contributions ( HV$_{GB}$: contribution of GBs, HV$_{part}$: contribution of particles, HV$_{ss}$: contribution of Fe in solid solution, HV$_{dis}$: contribution of dislocations deducted from the subtraction between HV$_{measured}$ and the different contributions, see text for details.

## 5. Conclusions

In this study, a thermally stable UFG structure of an Al-2%Fe alloy was successfully obtained using severe plastic deformation by HPT

1- UFG structure with 150 nm average grain size was obtained after 100 revolutions by HPT. This structure leads to an increase of the microhardness from 41±3 HV (as-cast) to 178±14 HV. After 1000 revolutions, a 140 nm average grain size was reached, giving rise to a microhardness of 200±22 HV.

2- Simultaneously to grain refinement, Al$_6$Fe brittle intermetallic particles have been progressively fragmented, dispersed and even partly dissolved, creating a supersaturated solid solution of Fe in the Al matrix (ranging from 0.36 to 0.89 at. % after 1000 revolutions by HPT). Averages particles sizes reach 70±4 nm and 40±7 nm after *N*=100 and *N*=1000 revolutions respectively.

3- These nanoscale particles, combined with the submicrometer grain size, dislocations, and Fe in solid solution strongly affect the microhardness. The corresponding contributions have been evaluated: Grain boundaries and dislocations each account for about 40 %, while nanoscale intermetallic particles and/or solid solution account for the rest.

4- Aging of these UFG structures leads to a limited growth of $Al_6Fe$ particles and heterogeneous precipitation of $Al_{13}Fe_4$ on defects. Estimations of the mobility of Fe atoms demonstrate that the diffusion and thus the precipitation kinetics is strongly promoted by strain-induced crystalline defects such as dislocations and GBs.

5- In any case, the nanoscaled intermetallic particles homogeneously introduced within the UFG structure strongly inhibits aluminum grain growth. The most uniform distribution of nanoscaled particles ($N$=1000) allows keeping an average grain size of about 300 nm only after 1h at 250°C. The comparison with theoretical estimates demonstrated that the structure is in a state nearly stabilized by intermetallic nanoparticles.

**Acknowledgements**

TEM *in-situ* experiments have been carried out on the GENESIS facility which is supported by the Région Normandie, the Métropole Rouen Normandie, the CNRS via LABEX EMC3 and the French National Research Agency as a part of the program

"Investissements d'avenir" with the reference ANR-11-EQPX-0020. The TEM-ACOM measurements were performed within the framework of the Centre of Excellence of Multifunctional Architectured Materials "CEMAM" n° ANR-10-LABX-44-01. Both the Agence Nationale de la Recherche (ANR) and the joint CNRS-JSPS 2017 summer program for the support to the trans-national collaboration are gratefully acknowledged for financial support (PRASA project - ANR-15-CE08-0029). This work was also supported in part by Grant-in-Aid for Scientific Research (S) from the MEXT, Japan (No. 26220909). HPT was carried out in the International Research Center on Giant Straining for Advanced Materials (IRC-GSAM) at Kyushu University.

GENESIS is supported by the Région Haute-Normandie, the Métropole Rouen Normandie, the CNRS via LABEX EMC and the French National Research Agency as a part of the program "Investissements d'avenir" with the reference ANR-11-EQPX-0020.

**References**
[1] O. Andreau, J. Gubicza, N. Xian Zhang, Y. Huang, P. Jenei, T.G. Langdon, Effect of short-term annealing on the microstructures and flow properties of an Al–1% Mg alloy processed by high-pressure torsion, Mater. Sci. Eng. A. 615 (2014) 231–239. doi:10.1016/j.msea.2014.07.018.
[2] C. Xu, M. Furukawa, Z. Horita, T.G. Langdon, Using ECAP to achieve grain refinement, precipitate fragmentation and high strain rate superplasticity in a spray-cast aluminum alloy, Acta Mater. 51 (2003) 6139–6149. doi:10.1016/S1359-6454(03)00433-6.
[3] Y. Estrin, A. Vinogradov, Extreme grain refinement by severe plastic deformation: A wealth of challenging science, Acta Mater. 61 (2013) 782–817. doi:10.1016/j.actamat.2012.10.038.
[4] D.C. Patil, K. Venkateswarlu, S.A. Kori, G. Das, M. Das, S.N. Alhajeri, T.G. Langdon, Mechanical property evaluation of an Al-2024 alloy subjected to HPT processing, IOP Conf. Ser. Mater. Sci. Eng. 63 (2014) 012085. doi:10.1088/1757-899X/63/1/012085.
[5] J.M. Cubero-Sesin, M. Arita, Z. Horita, High Strength and Electrical Conductivity of Al-Fe Alloys Produced by Synergistic Combination of High-Pressure Torsion and Aging, Adv. Eng. Mater. 17 (2015) 1792–1803. doi:10.1002/adem.201500103.
[6] J.M. Cubero-Sesin, H. In, M. Arita, H. Iwaoka, Z. Horita, High-pressure torsion for fabrication of high-strength and high-electrical conductivity Al micro-wires, J. Mater. Sci. 49 (2014) 6550–6557. doi:10.1007/s10853-014-8240-1.

[7]  J.M. Cubero-Sesin, M. Arita, M. Watanabe, Z. Horita, High strength and high electrical conductivity of UFG Al-2%Fe alloy achieved by high-pressure torsion and aging, IOP Conf. Ser. Mater. Sci. Eng. 63 (2014) 012117. doi:10.1088/1757-899X/63/1/012117.

[8]  R.K. Islamgaliev, K.M. Nesterov, J. Bourgon, Y. Champion, R.Z. Valiev, Nanostructured Cu-Cr alloy with high strength and electrical conductivity, J. Appl. Phys. 115 (2014) 194301. doi:10.1063/1.4874655.

[9]  X. Sauvage, E.V. Bobruk, M.Y. Murashkin, Y. Nasedkina, N.A. Enikeev, R.Z. Valiev, Optimization of electrical conductivity and strength combination by structure design at the nanoscale in Al–Mg–Si alloys, Acta Mater. 98 (2015) 355–366. doi:10.1016/j.actamat.2015.07.039.

[10] M.Y. Murashkin, I. Sabirov, A.E. Medvedev, N.A. Enikeev, W. Lefebvre, R.Z. Valiev, X. Sauvage, Mechanical and electrical properties of an ultrafine grained Al–8.5wt. % RE (RE=5.4wt.% Ce, 3.1wt.% La) alloy processed by severe plastic deformation, Mater. Des. 90 (2016) 433–442. doi:10.1016/j.matdes.2015.10.163.

[11] S.-S. Wang, J.-T. Jiang, G.-H. Fan, A.M. Panindre, G.S. Frankel, L. Zhen, Accelerated precipitation and growth of phases in an Al-Zn-Mg-Cu alloy processed by surface abrasion, Acta Mater. 131 (2017) 233–245. doi:10.1016/j.actamat.2017.03.074.

[12] A. Deschamps, G. Fribourg, Y. Bréchet, J.L. Chemin, C.R. Hutchinson, In situ evaluation of dynamic precipitation during plastic straining of an Al–Zn–Mg–Cu alloy, Acta Mater. 60 (2012) 1905–1916. doi:10.1016/j.actamat.2012.01.002.

[13] T. Chookajorn, H.A. Murdoch, C.A. Schuh, Design of Stable Nanocrystalline Alloys, Science. 337 (2012) 951–954. doi:10.1126/science.1224737.

[14] M. Saber, C.C. Koch, R.O. Scattergood, Thermodynamic Grain Size Stabilization Models: An Overview, Mater. Res. Lett. 3 (2015) 65–75. doi:10.1080/21663831.2014.997894.

[15] J.R. Trelewicz, C.A. Schuh, Grain boundary segregation and thermodynamically stable binary nanocrystalline alloys, Phys. Rev. B. 79 (2009) 094112. doi:10.1103/PhysRevB.79.094112.

[16] F. Liu, R. Kirchheim, Comparison between kinetic and thermodynamic effects on grain growth, Thin Solid Films. 466 (2004) 108–113. doi:10.1016/j.tsf.2004.03.018.

[17] F. Khan MD, S.K. Panigrahi, Achieving excellent thermal stability and very high activation energy in an ultrafine-grained magnesium silver rare earth alloy prepared by friction stir processing, Mater. Sci. Eng. A. 675 (2016) 338–344. doi:10.1016/j.msea.2016.08.077.

[18] H. Buken, E. Kozeschnik, A Model for Static Recrystallization with Simultaneous Precipitation and Solute Drag, Metall. Mater. Trans. A. 48 (2017) 2812–2818. doi:10.1007/s11661-016-3524-5.

[19] N. Wang, Y. Ji, Y. Wang, Y. Wen, L.-Q. Chen, Two modes of grain boundary pinning by coherent precipitates, Acta Mater. 135 (2017) 226–232. doi:10.1016/j.actamat.2017.06.031.

[20] R.K. Koju, K.A. Darling, L.J. Kecskes, Y. Mishin, Zener Pinning of Grain Boundaries and Structural Stability of Immiscible Alloys, JOM. 68 (2016) 1596–1604. doi:10.1007/s11837-016-1899-9.

[21] X. Boulnat, N. Sallez, M. Dadé, A. Borbély, J.-L. Béchade, Y. de Carlan, J. Malaplate, Y. Bréchet, F. de Geuser, A. Deschamps, P. Donnadieu, D. Fabrègue,

M. Perez, Influence of oxide volume fraction on abnormal growth of nanostructured ferritic steels during non-isothermal treatments: An in situ study, Acta Mater. 97 (2015) 124–130. doi:10.1016/j.actamat.2015.07.005.

[22] C.S. Smith, Grains, phases, and interphases: an interpretation of microstructure, Trans Met. Soc AIME Vol 175 P 15-51 1948. 175 (1948) 15–51.

[23] T. Gladman, Second phase particle distribution and secondary recrystallisation, Scr. Metall. Mater. 27 (1992) 1569–1573. doi:10.1016/0956-716X(92)90146-6.

[24] Y. Huang, J.D. Robson, P.B. Prangnell, The formation of nanograin structures and accelerated room-temperature theta precipitation in a severely deformed Al-4 wt.% Cu alloy, Acta Mater. 58 (2010) 1643–1657. doi:10.1016/j.actamat.2009.11.008.

[25] N. Hu, X.C. Xu, Influence of Dissolved Precipitated Phases on Mechanical Properties of Severely Deformed Al-4 Wt% Cu Alloys, Mater. Sci. Forum. 667–669 (2011) 1021–1026. doi:10.4028/www.scientific.net/MSF.667-669.1021.

[26] J.M. Cubero-Sesin, Z. Horita, Mechanical Properties and Microstructures of Al-Fe Alloys Processed by High-Pressure Torsion, Metall. Mater. Trans. A. 43 (2012) 5182–5192. doi:10.1007/s11661-012-1341-z.

[27] O.N. Senkov, F.H. Froes, V.V. Stolyarov, R.Z. Valiev, J. Liu, Microstructure and microhardness of an Al-Fe alloy subjected to severe plastic deformation and aging, Nanostructured Mater. 10 (1998) 691–698. doi:10.1016/S0965-9773(98)00107-X.

[28] O.N. Senkov, F.H. Froes, V.V. Stolyarov, R.Z. Valiev, J. Liu, Microstructure of Aluminum-Iron Alloys Subjected to Severe Plastic Deformation, Scr. Mater. 38 (1998). doi:10.1016/S1359-6462(98)00073-6.

[29] J.M. Cubero-Sesin, Z. Horita, Age Hardening in Ultrafine-Grained Al-2 Pct Fe Alloy Processed by High-Pressure Torsion, Metall. Mater. Trans. A. 46 (2015) 2614–2624. doi:10.1007/s11661-015-2876-6.

[30] B.D. Saller, G. Sha, L.M. Yang, F. Liu, S.P. Ringer, J.M. Schoenung, Iron in solution with aluminum matrix after non-equilibrium processing: an atom probe tomography study, Philos. Mag. Lett. 97 (2017) 118–124. doi:10.1080/09500839.2017.1292055.

[31] V.V. Tcherdyntsev, S.D. Kaloshkin, D.V. Gunderov, E.A. Afonina, I.G. Brodova, V.V. Stolyarov, Y.V. Baldokhin, E.V. Shelekhov, I.A. Tomilin, Phase composition and microhardness of rapidly quenched Al–Fe alloys after high pressure torsion deformation, Mater. Sci. Eng. A. 375–377 (2004) 888–893. doi:10.1016/j.msea.2003.10.207.

[32] N.A. Belov, A.A. Aksenov, D.G. Eskin, Iron in Aluminium Alloys: Impurity and Alloying Element, CRC Press, 2002.

[33] K. Liu, X. Cao, X.-G. Chen, A New Iron-Rich Intermetallic-AlmFe Phase in Al-4.6Cu-0.5Fe Cast Alloy, Metall. Mater. Trans. A. 43 (2012) 1097–1101. doi:10.1007/s11661-011-1004-5.

[34] D.L. Beke, I. Gödény, I.A. Szabó, G. Erdélyi, F.J. Kedves, On the diffusion of 59Fe into aluminium and Al Mn solid solutions, Philos. Mag. A. 55 (1987) 425–443. doi:10.1080/01418618708209907.

[35] Y. Du, Y.A. Chang, B. Huang, W. Gong, Z. Jin, H. Xu, Z. Yuan, Y. Liu, Y. He, F.-Y. Xie, Diffusion coefficients of some solutes in fcc and liquid Al: critical evaluation and correlation, Mater. Sci. Eng. A. 363 (2003) 140–151. doi:10.1016/S0921-5093(03)00624-5.

[36] V.V. Stolyarov, R. Lapovok, I.G. Brodova, P.F. Thomson, Ultrafine-grained Al–5 wt.% Fe alloy processed by ECAP with backpressure, Mater. Sci. Eng. A. 357 (2003) 159–167. doi:10.1016/S0921-5093(03)00215-6.

[37] K. Edalati, R. Miresmaeili, Z. Horita, H. Kanayama, R. Pippan, Significance of temperature increase in processing by high-pressure torsion, Mater. Sci. Eng. A. 528 (2011) 7301–7305. doi:10.1016/j.msea.2011.06.031.

[38] E. Rauch, M. Véron, S. Nicolopoulos, D. Bultreys, Orientation and Phase Mapping in TEM Microscopy (EBSD-TEM Like): Applications to Materials Science, Solid State Phenom. 186 (2012) 13–15. doi:10.4028/www.scientific.net/SSP.186.13.

[39] E. Rauch, S. Rouvimov, S. Nicolopoulos, P. Moeck, High Throughput Automated Crystal Orientation and Phase Mapping of Nanoparticles from HREM - TEM Images, Microsc. Microanal. 15 (2009) 756–757. doi:10.1017/S1431927609099607.

[40] D. Viladot, M. Véron, M. Gemmi, F. Peiró, J. Portillo, S. Estradé, J. Mendoza, N. Llorca-Isern, S. Nicolopoulos, Orientation and phase mapping in the transmission electron microscope using precession-assisted diffraction spot recognition: state-of-the-art results, J. Microsc. 252 (2013) 23–34. doi:10.1111/jmi.12065.

[41] K. Edalati, Z. Horita, A review on high-pressure torsion (HPT) from 1935 to 1988, Mater. Sci. Eng. A. 652 (2016) 325–352. doi:10.1016/j.msea.2015.11.074.

[42] W. Lefebvre-Ulrikson, Atom probe tomography: put theory into practice, 1st edition, London : Academic Press, 2016.

[43] I.R. Hughes, H. Jones, Coupled eutectic growth in Al-Fe alloys, J. Mater. Sci. 11 (1976) 1781–1793. doi:10.1007/BF00708256.

[44] A. Yamamoto, T. Kato, H. Tsubakino, Precipitation in an Al-300 ppm Fe Alloy, Mater. Trans. 45 (2004) 3106–3113. doi:10.2320/matertrans.45.3106.

[45] J.M. Cubero-Sesin, Z. Horita, Strengthening of Al through addition of Fe and by processing with high-pressure torsion, J. Mater. Sci. 48 (2012) 4713–4722. doi:10.1007/s10853-012-6935-8.

[46] B.L. Silva, A. Garcia, J.E. Spinelli, The effects of microstructure and intermetallic phases of directionally solidified Al–Fe alloys on microhardness, Mater. Lett. 89 (2012) 291–295. doi:10.1016/j.matlet.2012.08.130.

[47] K. Edalati, Z. Horita, Correlation of Physical Parameters with Steady-State Hardness of Pure Metals Processed by High-Pressure Torsion, Mater. Sci. Forum. 667–669 (2011) 683–688. doi:10.4028/www.scientific.net/MSF.667-669.683.

[48] M. Kawasaki, H.-J. Lee, B. Ahn, A.P. Zhilyaev, T.G. Langdon, Evolution of hardness in ultrafine-grained metals processed by high-pressure torsion, J. Mater. Res. Technol. 3 (2014) 311–318. doi:10.1016/j.jmrt.2014.06.002.

[49] A.P. Zhilyaev, K. Oh-ishi, T.G. Langdon, T.R. McNelley, Microstructural evolution in commercial purity aluminum during high-pressure torsion, Mater. Sci. Eng. A. 410 (2005) 277–280. doi:10.1016/j.msea.2005.08.044.

[50] Z. Horita, T.G. Langdon, Microstructures and microhardness of an aluminum alloy and pure copper after processing by high-pressure torsion, Mater. Sci. Eng. A. 410–411 (2005) 422–425. doi:10.1016/j.msea.2005.08.133.

[51] C. Xu, Z. Horita, T.G. Langdon, The evolution of homogeneity in an aluminum alloy processed using high-pressure torsion, Acta Mater. 56 (2008) 5168–5176. doi:10.1016/j.actamat.2008.06.036.

[52] C. Xu, T.G. Langdon, Three-dimensional representations of hardness distributions after processing by high-pressure torsion, Mater. Sci. Eng. A. 503 (2009) 71–74. doi:10.1016/j.msea.2008.04.083.

[53] V.V. Tcherdyntsev, S.D. Kaloshkin, D.V. Gunderov, E.A. Afonina, I.G. Brodova, V.V. Stolyarov, Y.V. Baldokhin, E.V. Shelekhov, I.A. Tomilin, Phase composition and microhardness of rapidly quenched Al–Fe alloys after high pressure torsion deformation, Mater. Sci. Eng. A. 375–377 (2004) 888–893. doi:10.1016/j.msea.2003.10.207.

[54] Kattner, Binary Alloy Phase Diagrams, T.B. Massalski, 1986.

[55] American Society for Metals, K.R. Van Horn, Aluminium. volume 1, volume 1, American Society for Metals, Ohio, 1967.

[56] D. Akama, Z. Horita, K. Matsuda, S. Hirosawa, Aging Behavior of Al-Mg-Si Alloys Processed by High-Pressure Torsion, Mater. Sci. Forum. 667–669 (2011) 259–264. doi:10.4028/www.scientific.net/MSF.667-669.259.

[57] E. Bagherpour, M. Reihanian, H. Miyamoto, Tailoring particle distribution non-uniformity and grain refinement in nanostructured metal matrix composites fabricated by severe plastic deformation (SPD): a correlation with flow stress, J. Mater. Sci. 52 (2017) 3436–3446. doi:10.1007/s10853-016-0632-y.

[58] X. Sauvage, P. Jessner, F. Vurpillot, R. Pippan, Nanostructure and properties of a Cu–Cr composite processed by severe plastic deformation, Scr. Mater. 58 (2008) 1125–1128.

[59] X. Quelennec, A. Menand, J.M.L. Breton, R. Pippan, X. Sauvage, Homogeneous Cu–Fe supersaturated solid solutions prepared by severe plastic deformation, Philos. Mag. 90 (2010) 1179–1195. doi:10.1080/14786430903313682.

[60] K.S. Kormout, R. Pippan, A. Bachmaier, Deformation-Induced Supersaturation in Immiscible Material Systems during High-Pressure Torsion, Adv. Eng. Mater. 19 (n.d.) 1600675. doi:10.1002/adem.201600675.

[61] G. RUMMEL, T. Zumkley, M. Eggersmann, K. Freitag, H. Mehrer, Z. Metallkd, Diffusion of implanted 3d-transition elements in aluminum, Part I, Temperature Dependence, Z Met. 85 (1995) 122–130.

[62] G.M. Hood, The diffusion of iron in aluminium, Philos. Mag. J. Theor. Exp. Appl. Phys. 21 (1970) 305–328. doi:10.1080/14786437008238419.

[63] W.B. Alexander, L.M. Slifkin, Diffusion of Solutes in Aluminum and Dilute Aluminum Alloys, Phys. Rev. B. 1 (1970) 3274–3282. doi:10.1103/PhysRevB.1.3274.

[64] S. Mantl, W. Petry, K. Schroeder, G. Vogl, Diffusion of iron in aluminum studied by M\"ossbauer spectroscopy, Phys. Rev. B. 27 (1983) 5313–5331. doi:10.1103/PhysRevB.27.5313.

[65] Y. Du, Y.A. Chang, B. Huang, W. Gong, Z. Jin, H. Xu, Z. Yuan, Y. Liu, Y. He, F.-Y. Xie, Diffusion coefficients of some solutes in fcc and liquid Al: critical evaluation and correlation, Mater Sci Eng A. 363 (2003) 140–151. doi:10.1016/S0921-5093(03)00624-5.

[66] A. Einstein, Investigations on the Theory of the Brownian Movement, edited with notes by R. Fürth, Dover publications, 1956.

[67] L. Shaw, H. Luo, J. Villegas, D. Miracle, Thermal stability of nanostructured Al93Fe3Cr2Ti2 alloys prepared via mechanical alloying, Acta Mater. 51 (2003) 2647–2663. doi:10.1016/S1359-6454(03)00075-2.

[68] R. Elst, J. van Humbeeck, L. Delaey, Evaluation of grain growth criteria in particle-containing materials, Acta Metall. 36 (1988) 1723–1729. doi:10.1016/0001-6160(88)90240-4.

[69] T. Gladman, P. D, On the theory of the effect of precipitate particles on grain growth in metals, Proc R Soc Lond A. 294 (1966) 298–309. doi:10.1098/rspa.1966.0208.

[70] M. Hillert, Inhibition of grain growth by second-phase particles, Acta Metall. 36 (1988) 3177–3181. doi:10.1016/0001-6160(88)90053-3.

[71] S.A. Humphry-Baker, C.A. Schuh, Suppression of grain growth in nanocrystalline $Bi_2Te_3$ through oxide particle dispersions, J. Appl. Phys. 116 (2014) 173505. doi:10.1063/1.4901235.

[72] A.P. Zhilyaev, T.G. Langdon, Using high-pressure torsion for metal processing: Fundamentals and applications, Prog. Mater. Sci. 53 (2008) 893–979. doi:10.1016/j.pmatsci.2008.03.002.

[73] G.J. Mahon, G.J. Marshall, Microstructure-property relationships in O-temper foil alloys, JOM. 48 (1996) 39–42. doi:10.1007/BF03222965.

[74] R. Labusch, A Statistical Theory of Solid Solution Hardening, Phys. Status Solidi B. 41 (1970) 659–669. doi:10.1002/pssb.19700410221.

[75] B.Q. Han, F.A. Mohamed, E.J. Lavernia, Tensile behavior of bulk nanostructured and ultrafine grained aluminum alloys, J. Mater. Sci. 38 (2003) 3319–3324. doi:10.1023/A:1025150407956.

[76] J.W. Martin, Precipitation Hardening, Second Edition, Butterworth-Heinemann, 2012.

[77] E. Orowan, Discussion on internal stresses. Symposium on Internal Stresses in Metals and Alloys, (1948) 451–453.

[78] E.O. Hall, The Deformation and Ageing of Mild Steel: III Discussion of Results, Proc. Phys. Soc. Sect. B. 64 (1951) 747. doi:10.1088/0370-1301/64/9/303.

[79] N.J. Petch, The Cleavage Strength of Polycrystals, J. Iron Steel Inst. 174 (1953) 25–28.

[80] T. Shanmugasundaram, M. Heilmaier, B.S. Murty, V.S. Sarma, On the Hall–Petch relationship in a nanostructured Al–Cu alloy, Mater. Sci. Eng. A. 527 (2010) 7821–7825. doi:10.1016/j.msea.2010.08.070.

[81] J.E. Bailey, P.B. Hirsch, The dislocation distribution, flow stress, and stored energy in cold-worked polycrystalline silver, Philos. Mag. J. Theor. Exp. Appl. Phys. 5 (1960) 485–497. doi:10.1080/14786436008238300.

[82] K. Ma, H. Wen, T. Hu, T.D. Topping, D. Isheim, D.N. Seidman, E.J. Lavernia, J.M. Schoenung, Mechanical behavior and strengthening mechanisms in ultrafine grain precipitation-strengthened aluminum alloy, Acta Mater. 62 (2014) 141–155. doi:10.1016/j.actamat.2013.09.042.

[83] T.H. Courtney, Mechanical Behavior of Materials: Second Edition, Waveland Press, 2005.

[84] Z. Horita, K. Ohashi, T. Fujita, K. Kaneko, T.G. Langdon, Achieving High Strength and High Ductility in Precipitation-Hardened Alloys, Adv. Mater. 17 (2005) 1599–1602. doi:10.1002/adma.200500069.

[85] S.S. Nayak, M. Wollgarten, J. Banhart, S.K. Pabi, B.S. Murty, Nanocomposites and an extremely hard nanocrystalline intermetallic of Al–Fe alloys prepared by mechanical alloying, Mater. Sci. Eng. A. 527 (2010) 2370–2378. doi:10.1016/j.msea.2009.12.044.

[86] C. Gammer, H.P. Karnthaler, C. Rentenberger, Unexpected grain size reduction by heating in bulk nanocrystalline FeAl, J. Alloys Compd. 633 (2015) 384–389. doi:10.1016/j.jallcom.2015.01.186.

[87] T.S. Orlova, A.M. Mavlyutov, A.S. Bondarenko, I.A. Kasatkin, M.Y. Murashkin, R.Z. Valiev, Influence of grain boundary state on electrical resistivity of ultrafine grained aluminium, Philos. Mag. 96 (2016) 2429–2444. doi:10.1080/14786435.2016.1204022.

[88] X.K. Zhu, Y.J. Chao, Effects of temperature-dependent material properties on welding simulation, Comput. Struct. 80 (2002) 967–976. doi:10.1016/S0045-7949(02)00040-8.

[89] J.S. Robinson, W. Redington, The influence of alloy composition on residual stresses in heat treated aluminium alloys, Mater. Charact. 105 (2015) 47–55. doi:10.1016/j.matchar.2015.04.017.